\documentclass[twocolumn,showpacs,amsmath,amssymb,pra,floatfix]{revtex4}

\usepackage{bm} 
\usepackage{epsf} 
\usepackage[english]{babel} 
\usepackage{graphicx} 
\def\pb{$\bar{p}$} 
\def\mm{$\mu {\, \bar{}}$}
\def\spb{$\textstyle S^{ \bar{p}}$} 
\def\smm{$\textstyle S^{\mu {\, \bar{}}}$}
\def\ssin{$\textstyle S^{\rm \,sin}$}
\def\ae#1{$\bar{\epsilon}^{#1}$}%
\def\aei#1{$\bar{\epsilon}_{\rm ion}^{#1}$}%
\def\aee#1{$\bar{\epsilon}_{\rm exc}^{#1}$}%
\def\emax{E_{\rm max}}%
\def\hmol{H$_2$}%
\def\h2p{H$_{2}^{+}$}%

\def\opp#1{\hat{#1}}

\newcommand{\diff}[1]{\textrm{d}#1}

\newcommand{\mean}[1]{\left\langle\,#1\,\right\rangle}

\begin{document}

\bibliographystyle{apsrev} 

%
%
 
\title{Stopping power of antiprotons in H, H${\bf _2}$, and He
  targets}     
 
\author{Armin L\"uhr} 
\author{Alejandro Saenz} 
 
\affiliation 
{Institut f\"ur Physik,  
AG Moderne Optik, Humboldt-Universit\"at zu Berlin, Hausvogteiplatz 5-7, 
D-10117 Berlin, Germany.}  

\date{\today} 
 
             
\begin{abstract}\label{txt:abstract} 
The stopping power of antiprotons in atomic and molecular hydrogen as well as
helium was calculated in an impact-energy range from 1 keV to 6.4 MeV. In the
case of \hmol\ and He the targets were described with a single-active electron
model centered on the target. The collision process was treated with the
close-coupling formulation of the impact-parameter method.
An extensive comparison of the present results with theoretical and experimental
literature data was performed in order to evaluate which of the partly
disagreeing theoretical and experimental data are most reliable.
Furthermore, 
the size of the corrections to the first-order stopping number, the average
energy transferred to the target electrons, and the relative importance of the
excitation and the ionization process for the energy loss of the
projectile was determined. Finally, the stopping power of the H, \hmol , and
He targets were directly compared revealing specific similarities and
differences of the three targets.  
\end{abstract}

\maketitle


\section{Introduction} 
\label{sec:introduction}
During the last two decades low-energy antiproton (\pb ) collisions have
evolved from an exotic system into a powerful tool to achieve understanding of
fundamental processes in atoms, molecules, and solids.  
Obvious advantages of \pb\ are that they have a negative charge
and are heavy in comparison with an electron and thus ideal projectiles from a
theoretical point of view.  
A number of theoretical efforts have been done, e.g., for low-energy
collisions with He atoms focusing on the single and double ionization cross
sections. They were stimulated by discrepancies between experiment and theory
lasting for more than a decade which only recently could partly be resolved
~\cite{anti:knud08}. In the case of  
ionization and excitation of the simplest two-electron
molecule  \hmol\ by \pb\ impact the rather sparse information
\cite{anti:hvel94,anti:ermo93} could be extended recently
\cite{anti:luhr08a}. Precise data on   
\pb\ +\,\hmol\ are, however, of great interest in many fields. 
They can also be used to determine the stopping power which is needed in
several applications. It is a prerequisite for the
design of low-energy \pb\ storage rings taking the interactions with
residual-gas atoms and molecules into account. 
But also the maximum of the stopping power is of
importance for the preparation of accurate (future) experiments with
low-energy \pb\ which are dealing with, e.g., 
antiprotonic atoms and therefore the capture and annihilation process of \pb ,
inelastic scattering events, or the formation of antihydrogen. 
These experiments are intended to shed more light on fundamental
questions regarding the matter-antimatter interaction like tests of the CPT
invariance and measurements of the gravity of antimatter. 
The outcome of collision experiments with \pb\ can in turn be
used as a stringent test of competing theoretical approaches.

The quantum mechanical formulation of the energy loss of fast charged
particles in matter is based on the theory by Bethe
\cite{sct:beth30,sct:beth32}. He derived the stopping power in the first-order
Born approximation which is proportional to the projectile charge squared
$Z_p^2$. In Bethe's model, the stopping power $-dE/dx$ or energy loss per unit
length of a charged particle with the velocity $v$ can be written as  
\begin{equation}
  \label{eq:stopping}
  -\frac{dE}{dx} = NS(v) = N \frac{4\pi e^4 Z}{m}\frac{Z_p^2}{v^2} L(v)\,,
\end{equation}
where $N$ is the density of atoms of atomic number $Z$ in the stopping medium,
$m$ is the electron mass, and $e$ is the elementary charge. $S(v)$ is the
stopping cross section (related to the stopping power by $N$) and $L(v)$ is the
velocity-dependent stopping number.  

While Eq.\ (\ref{eq:stopping}) which is quadratic in $Z_p$ works sufficiently
well for 
high non-relativistic velocities it was a surprise when it was found in an
experiment that the range of negative pions was longer
than that of positive pions of equal momentum. The existence of this
phenomenon was 
later fully confirmed with negative and positive hyperons by Barkas {\it
  et al.}\ \cite{sct:bark63}. This so-called 
Barkas effect has been interpreted as a polarization effect in the stopping
material depending on the charge of the projectile. It appears 
as the second term in the implied Born expansion of the energy loss and is
proportional to $Z_p^3$. Following Lindhard \cite{sct:lind76}, the stopping
number may be expanded in a Born series in $Z_p$ as 
\begin{equation}
  \label{eq:born_series}
  L(v) = \sum_{i=0}^{\infty} Z_p^i L_i(v)\,.
\end{equation}
where $L_0$ ($S\propto Z_p^2$) is the Bethe term. The second term $L_1$
($S\propto Z_p^3$) also referred to as Barkas correction is the first
odd-order term in the Born series and reflects the asymmetry of the energy loss
between charge conjugated particles.

With the advent of the Low-Energy Antiproton Ring (LEAR) at CERN, \pb\
beams with improved quality at low energy became available, making an accurate
comparison of stopping powers for antiprotons and protons ($p$)
feasible. The first measurements were performed for solid silicon
\cite{anti:ande89a}. The \pb\ stopping powers \spb\ for various solid targets
which where obtained in more recent experiments \cite{anti:moll02} at the
Antiproton Decelerator (AD) 
were found to be smaller by 35-55\% than those for $p$ collisions and
confirmed therefore an asymmetry between charge conjugated projectiles. 
These measurements also strongly supported a proportionality of the
stopping power to the velocity  below the stopping maximum expected for 
a point-like projectile.

Stopping powers for \pb\ in \hmol\ and He were measured by the OBELIX
Collaboration 
\cite{anti:adam93,anti:agne95} also at LEAR for a kinetic energy range of the
\pb\ from about 0.5 keV to 1.1 MeV. In these experiments, a focus was put on
the investigation 
of the Barkas effect. Their results indicate fundamental differences --- calling
for a thorough investigation of the involved stopping mechanism --- between
\pb\ stoppings in the simplest gases (He, \hmol ) and in solid targets below 
some MeV \cite{anti:agne95,anti:lodi02,anti:lodi04}. Particularly, below the 
\pb\ stopping-power maximum no velocity proportionality could be
observed. Above the maximum the stopping power \spb\ for \pb\ collisions was
claimed to be even larger than for $p$ impact ($S^p$) with a difference
$S^{\bar{p}} - S^p$ of 21\% $\pm$3\%  and 15\% $\pm$5\% around a kinetic
energy of 600 keV for \hmol\ \cite{anti:lodi02} and 700 keV for He
\cite{anti:lodi04}, respectively. 
In a very recent effort \cite{anti:bian08} the measured He data
\cite{anti:agne95} were reconsidered. After an extended analysis of the data
it was claimed that a part of the antiprotons have to be reflected by the wall
of the gas vessel in order to bring the simulated results in accordance with
the experimentally measured data. A sizeable influence of this newly considered
reflection process on the previously analyzed stopping power is, however, not
expected by these authors \cite{anti:lodi08}. Although the data were taken
more than a decade ago theoretical investigations have not been able to fully
reproduce the experimental findings concerning the slowing down of the
antiprotons; especially for \hmol\ targets.   

Approximately at the same time experiments for negatively charged muons (\mm\,)
stopping in \hmol\ and He gases were performed at the PSI
\cite{sct:kott87,sct:haus93,sct:kott94}. In these experiments basically the
excitation cross sections were determined by measuring the time-distribution
of the scintillation light emitted from the excited targets during the slowing
down of the projectile. In order to derive the \smm\ also experimental \pb\
ionization cross sections and experimental and theoretical data for the mean
energy transfer for ionization and excitation of the target were used. In
contrast to the \pb\ results the \smm\ were found to stay below $S^p$ for
energies above the stopping maximum $E>\emax$. 
However, the analysis of the \mm\ {} data it was assumed 
that for fast particles with a velocity $v \ge 0.1\,c$ (corresponding to a
antiproton energy of approximately 4.7 MeV) the Bethe-Bloch stopping formula
is valid. In a more recent measurement for \mm\ in an \hmol\ gas target
performed by the same authors the stopping power was measured directly
\cite{anti:schm98}. The results also stay below the proton stopping power for
$E>\emax$. Although the uncertainties of the latter experiment are
considerably larger those in \cite{sct:kott87,sct:haus93,sct:kott94}
(and thus its results are not discussed quantitatively here)  
these uncertainties are caused by totally different systematic errors than in
the earlier muon experiments providing therefore results which are independent
of the earlier findings.   

Except for deviations at small projectile velocities $v$ the total \pb\ and
\mm\ stopping powers should be the same at a given $v$, \smm $(v)$ = \spb
$(v)$. The deviations among the experimental results are, however, of the
order of 20\% indicating the experimental difficulties and uncertainties. 

A peculiarity in the context of antiproton scattering and in particular for
the stopping power is the fact that in
the case of hydrogen targets all experiments were done for
\emph{molecules} while the theoretical description on the other hand
concentrates mainly on \emph{atomic} targets
\cite{anti:schi96,anti:cabr05,anti:cust05,anti:cohe83}. The evident deviations
between the theoretical atomic and experimental molecular hydrogen results for
\spb\ were therefore claimed to origin from molecular effects
\cite{anti:schi96,anti:cabr05,anti:cust05}. The naive picture of an \hmol\
molecule as being 
basically the same as two individual H atoms has been shown to be inadequate
for the type of collision processes considered here \cite{anti:luhr08a}. It is
one aim of the present work to treat the \emph{atomic} and \emph{molecular}
hydrogen targets 
separately in order to figure out the differences and also to compare directly
to the experimental findings. The \hmol\ molecule and the He atom are
described with an effective one-electron model potential which was discussed
in detail in \cite{dia:luhr08} and already applied for the determination of
ionization and excitation cross sections for \pb\ + \hmol\ collisions
\cite{anti:luhr08a}. 
Also, the incongruity among the experimental results is discussed in view
of the present findings. Possible deficiencies of the used model description in
connection to the stopping power are discussed using the He target which is
studied more rigorously theoretically as well as experimentally.  

The following section gives a short review on the coupled-channel method
applied to the energy-loss calculations and the employed model potential for
the target description. In Sec.\ \ref{sec:results} the present stopping powers
for H, \hmol , and He are presented and compared to literature. 
A more detailed discussion of the results follows in Sec.\
\ref{sec:discussion}. This includes the 
determination of the Barkas effect and the consideration of the discrepancies
among the stopping powers available in literature. Section
\ref{sec:conclusion} concludes on the findings and gives a short
outlook. Atomic units are used unless stated otherwise.

%
\section{Method} 
\label{sec:method} 
It can be assumed that the total stopping power of a heavy particle
consists of an electronic and a nuclear part. The nuclear stopping power
is of importance for very small impact velocities. For \pb\ collisions
with $E>10$ keV it is, however, fully dominated by the electronic stopping
power for hydrogen and helium targets
\cite{anti:bert96,anti:schi96,anti:cabr05,anti:cust05}. In what follows, only
the electronic part of the stopping power $S$ is determined. 

A natural approach to
measure the stopping power of a medium is to quantify the energy difference of
the projectiles before and behind the target medium of a certain thickness and
density which may be variable. Instead of looking at the energy which is lost
by the 
projectile it is on the other hand also possible to consider the energy gain
of the stopping medium due to the interaction with the projectile. Both
perspectives are equivalent since the sum of the energy loss by the projectile
and the energy gain by the medium has to be zero. In the present investigation
the latter point of view is used to determine the stopping power 
\begin{equation} 
  \label{eq:stopping_power} 
  S =  \sum_f (\epsilon_f - \epsilon_i)\,\sigma_{f} \,,   
\end{equation} 
where $\sigma_{f}$ is the cross section for a transition from the initial
state $i$ into a final state $f$. Accordingly, $\epsilon_i$ and $\epsilon_f$ are
the energies of the states $i$ and $f$, respectively. They express
the energy transfer from the projectile to the target needed for the
transition and therefore the energy which is lost by the projectile.  

In order to obtain the electronic stopping power a general,
non-perturbative method for calculating ion collisions is used which has
been implemented recently \cite{anti:luhr08,anti:luhr08a}. It is based on a
close-coupling approach within an atomic-orbital description of the electrons
of the individual target atoms in the stopping medium. An advantage of this
approach is the fact that within the space spanned by the basis functions used
for the expansion of the time-dependent scattering wave function the
projectile-target interaction is treated in infinitely high order. In recent
applications ionization and excitation cross sections as well as
electron-energy spectra were determined for antiproton and proton collisions 
with alkali-metal atoms \cite{anti:luhr08,anti:luhr09a} and molecular hydrogen
\cite{anti:luhr08a,dia:luhr08,anti:luhr09b}.   

The collision process is considered in a semi-classical way using the impact
parameter method. Thereby, the target electrons are treated
quantum mechanically while the heavy projectile moves on a straight classical
trajectory ${\bf R}(t)={\bf b} + {\bf v} t$ given by the impact parameter
${\bf b}$ and the velocity ${\bf v}$ which are parallel to the $x$ and $z$
axis, respectively, and $t$ is the time.
 
An effective one-electron description of the collision process is used,
\begin{equation} 
  \label{eq:tdSE} 
        i {\frac{\partial}{\partial t}} \Psi({\bf r},{\bf R}(t)) =  
        \left (  \opp{H}_0 + \opp{V}_{\rm int}({\bf r},{\bf R}(t)) \right )
         \Psi({\bf r},{\bf R}(t))\, , 
  \end{equation} 
where $\mathbf{r}$ is the electron coordinate and the interaction between the
projectile with charge $Z_p$ and the target electron is expressed by the
time-dependent interaction potential
\begin{equation} 
  \label{eq:interaction_potential} 
         \opp{V}_{\rm int}({\bf r},{\bf R}(t)) 
         = -\frac{Z_p}{\left| \mathbf{r-R}(t) \right|} 
\, . 
  \end{equation} 
The time-dependent scattering wave function 
\begin{equation} 
  \label{eq:psi}
  \Psi(r,{\bf R}(t))= \sum_j c_j({\bf R}(t))\,\phi_j(r)
\end{equation} 
is expanded in eigenstates $\phi_j$ of the time-independent target Hamiltonian
\begin{equation} 
  \label{eq:target_Hamiltonian}
  \opp{H}_0= -\frac{1}{2}\, \nabla^2 + \opp{V}_{\rm target}(r)\,.
\end{equation} 
The $\phi_j$ are centered on the target nucleus. Their radial part is
expanded in B-spline functions while their angular part is expressed in
spherical harmonics.  

The potential $V_{\rm target}$ in Eq.\ (\ref{eq:target_Hamiltonian})
\begin{equation} 
  \label{eq:target_potential} 
  V_{\rm target}(r) = - \frac{1}{r}\, \left(  
                    1 + \frac{\alpha}{|\alpha|}\,\exp\,\left[ 
                                        -\frac{2\,r}{|\alpha|^{1/2}} 
                                      \right ] 
                  \right)\,,  
\end{equation} 
used for the (effective) one-electron description of the target was proposed
in \cite{sfm:vann08} and discussed in detail in 
\cite{dia:luhr08}. 
The potential in Eq.\ (\ref{eq:target_potential}) becomes parameter free by
requiring $\alpha$ to be chosen in such a way that the
ionization potential of the model coincides with the one of the
target. Additionally, in the limit $\alpha \rightarrow 0$ one obtains the
potential for atomic hydrogen $V_{\rm target}(r) = V_{\rm  H}(r)=-1/r$ as well
as in the limit $r\rightarrow 0$ for arbitrary $\alpha$. 
A value of $\alpha=0.8791$ yields the 
correct ionization potential for ground state He atoms. In the case of an
\hmol\ molecule one has to keep in mind that the ionization potential is --- in
a fixed nuclei approximation --- dependent on the internuclear distance $R_n$
between the two nuclei. The relation between $R_n$ and $\alpha$ is given in
\cite{dia:luhr08}. 

It has been shown in \cite{anti:luhr08a} that within the Born-Oppenheimer
approximation cross sections 
for antiproton collisions with \hmol\ are linear in $R_n$ around $R_n=1.4$
a.u. This property was used according to \cite{nu:saen97b} in order to obtain
cross sections which are independent of $R_n$ and to a certain extent also
account for the motion of the \hmol\ nuclei. However, the rovibronic motion
is due to the use of closure energetically not resolved. 
The procedure used here employs closure, exploits the linear behavior in $R_n$
of the cross section around $R_n=1.4$ a.u., and finally performs the
calculations at $R_n=\mean{R_n}=1.4487$ a.u. Therefore, the value $\alpha =
0.13308$ is used in the present calculations which results in an ionization
potential of the model which is equal to the ionization potential of \hmol\
for $R_n= \mean{R_n} = 1.4487$ a.u.

The expansion of $\Psi(r,{\bf R}(t))$ in Eq.\ (\ref{eq:tdSE}) as given in Eq.\
(\ref{eq:psi}) leads to 
coupled, first-order differential equations for the expansion coefficients
$c_j({\bf R}(t))$ for every trajectory ${\bf R}(t)$, i.e., for every $v$ (and
therefore for every impact energy $E=(1/2)\,M_p\,v^2$ where $M_p$ is the
projectile mass) and $b$.  
The differential equations are integrated in a finite $z$ range $-40
\textrm{ a.u.} \le z=v t \le 70$ a.u.\ with the initial condition $c_j({\bf R}
(t_i$=$-40/v)) = \delta_{ji}$ that the target is initially in its ground
state $\phi_i$. 

The single-electron probability for a transition into the final
state $\phi_f$ at $t_f=70/v$ is given by
\begin{equation}
  \label{eq:se_probability}
  p_f (b,v) = | c_f(b,v,t_f)|^2\,
\end{equation}
and is used for H atoms.
In the case of the two-electron targets \hmol\ and He the independent particle
model (IPM) is employed. 
It assumes that both electrons have the same transition probabilities which
are simply given by the single-electron probabilities Eq.\
(\ref{eq:se_probability}). 
Furthermore, the electrons are considered as being independent of each other
in the way that both feel the same attractive potential $V_{\rm target}$ which
includes the interaction with the other electron only by an averaged screening
of the nuclear charge. 
As a consequence the total stopping power due
to one electron is ---in contrast to the cross sections obtained with the
IPM--- independent from and equal to the one of the other 
electron. Therefore, in the case of targets with $N$ electrons the final
stopping power given in Eq.\ (\ref{eq:stopping_power}) computed for a single
electron by using Eq.\ (\ref{eq:se_probability}) has to be multiplied with
the factor $N$ in order to sum up the contributions from all $N$ independent
electrons. 

This argument can also be expressed in a more formal way starting with the
relation 
\begin{equation}
  \label{eq:unity_relation}
  1 = \sum_j p_j = \sum_j p_j   \sum_k p_k =  \sum_{j,k} p_j p_k\,,
\end{equation}
where it has been used that the sum over all single-electron transition
probabilities $p_j$ including the probability for staying in the initial state
is unity. The indices $j$ and $k$ are meant to indicate  one of the
electrons. Actually, the transition probabilities all depend on $b$ and $v$
which is for the sake of clarity not explicitly written in this derivation. In
view of the stopping power (cf. Eq.\ (\ref{eq:stopping_power})) the transition
probabilities in Eq.\ (\ref{eq:unity_relation}) are multiplied with the sum of
energies $\tilde{\epsilon}_j + \tilde{\epsilon}_k$ needed for the transitions
of one electron into state $\phi_j$ and the other into $\phi_k$
%
\begin{eqnarray}
  \label{eq:de_probability}
  \sum_{j,k} p_j p_k (\tilde{\epsilon}_j + \tilde{\epsilon}_k)   &=& 
   \sum_{j,k} p_j p_k \,\tilde{\epsilon}_j + \sum_{j,k} p_j p_k
   \,\tilde{\epsilon}_k \\ 
  \label{eq:de_intermediate}
  &=&  2 \sum_{j,k} p_j p_k \,\tilde{\epsilon}_j  \\
  &=& 2 \sum_{j} p_j \,\tilde{\epsilon}_j \underbrace{\ \, \sum_{k}\, p_k\,
  }_{\mbox{= 1}}\\ 
  \label{eq:de_final}
  &=& 2 \sum_{j} p_j \,\tilde{\epsilon}_j 
      = \sum_{j} p_j\,(2\tilde{\epsilon}_j)
      \,.\quad
\end{eqnarray}
It should be noted that the use of the sum of single-electron energies for both
electrons is an approximation which seems, however, to be consistent within
the employed IPM. This approximation is reasonable if one-electron transitions
are dominating the electronic energy loss.
Finally, the last line can be interpreted in the way
that the total single-electron
stopping power is multiplied with a factor two which accounts for both
electrons as mentioned above. Note, that the sum of all probabilities is still
unity and is therefore conserved as it should be.

A similar derivation as in Eqs.\ (\ref{eq:de_probability}--\ref{eq:de_final})
can be used in the case that the summation runs only over a limited number of
final states. As an example the contribution to the stopping power due to double
ionization shall be considered. Then the indices $j$ and $k$  only take
continuum states into account leading to a restricted summation indicated
by an asterisk above the sum. Equations
(\ref{eq:de_intermediate}-\ref{eq:de_final}) then take the form
\begin{eqnarray}
  \label{eq:de_ionization}
  2 \sum_{j,k}^{*} p_j p_k \,\tilde{\epsilon}_j 
  &=& 2 \,\sum_{k}^{*}\, p_k\, \sum_{j}^{*} p_j \,\tilde{\epsilon}_j \\ 
  \label{eq:de_ionization_final}
  &=& 2 \quad\,  p_I\quad\, \sum_{j}^{*} p_j\,\tilde{\epsilon}_j\,, 
\end{eqnarray}
where $p_I$ is the sum of all single-electron transition probabilities into
continuum states. 
%
%

%
%
%
%

The contribution to the stopping power from all electron transitions into the
continuum is obtained by relaxing the restriction on the sum over $k$ in Eq.\
(\ref{eq:de_ionization}) to all possible final states of the other electron
\begin{equation}
  \label{eq:total_ionization}
    2\,\sum_{k}\sum_{j}^{*} p_j p_k\,\tilde{\epsilon}_j
  = 2\,\sum_{k}p_k\sum_{j}^{*} p_j \,\tilde{\epsilon}_j
  = 2\,\sum_{j}^{*} p_j\,\tilde{\epsilon}_j\,,
\end{equation}
which yields a factor one instead
of $p_I$ in  Eq.\ (\ref{eq:de_ionization_final}). 
Thus, the same result as in the one-electron case is obtained which is just
multiplied with a factor two. However, in the case of two-electron targets
this is the sum of contributions due to single and double ionization. 
Accordingly, the contribution to the stopping power from all electron
transitions into bound states is given by 
\begin{equation}
  \label{eq:total_excitation}
    2\,\sum_{k}\sum_{j}^{\_\_\_} p_j p_k\,\tilde{\epsilon}_j
  = 2\,\sum_{k}p_k\sum_{j}^{\_\_\_} p_j \,\tilde{\epsilon}_j
  = 2\,\sum_{j}^{\_\_\_} p_j\,\tilde{\epsilon}_j\,,
\end{equation}
%
where the bar above the sum indicates that the summation is restricted to
bound states only. The sum of the contributions due to ionization (Eq.\
(\ref{eq:total_ionization})) and excitation (Eq.\ (\ref{eq:total_excitation}))
obviously yields the correct total result as given in Eq.\ (\ref{eq:de_final}).



The results in Eqs.\ (\ref{eq:de_final}--\ref{eq:total_excitation})
for the stopping power seem to contradict with the way how two-electron cross
sections are extracted from single-electron probabilities employing the
IPM. For cross sections it is important that the probability for 
double excitation (or ionization) is only counted once and not twice in order
to preserve the sum of probabilities being unity. The
factor two in the case of the stopping power, however, appears
not due to an increase of the probability for double transitions but because
of the fact that in these transitions both electrons gain energy and the
probability has therefore to be weighted with the number of electrons by what
the seeming contradiction is resolved.

The cross section $\sigma_{f}$ required for the determination of the total
stopping power in Eq.\ (\ref{eq:stopping_power}) for a certain $v$ is
obtained (using the cylindrical symmetry of the collision system) by an
integration over $b$,      
\begin{equation} 
  \label{eq:cross_section} 
 \sigma_{f}(v) = 2\, \pi\, \int p_{f}(b,v) \,b\; 
                         \diff{b}\,,      
\end{equation} 
where 
$ p_f$ is given in Eq.\ (\ref{eq:se_probability}). In the case of two-electron
targets the result obtained with Eq.\ (\ref{eq:stopping_power}) has to be
multiplied ---in accordance with Eq.\ (\ref{eq:de_final})--- with the factor
two.   

The present results were calculated with a basis set similar to that used
in \cite{anti:luhr08} including orbitals with
angular momenta up to $l = 7$. An energy cutoff for the continuum states of
250 a.u.\ was used leading to about 260 B-spline functions per angular
momentum. A non-linear knot sequence was employed for the radial
coordinate. The interaction potential in
Eq.\ (\ref{eq:interaction_potential}) causes $l$ and $m_l$ mixing. In order to
reduce the numerical effort only orbitals with magnetic quantum numbers
$|m_l|\le 3$ were taken into account. 
Exploiting the symmetries of the collision system 
$\Psi$ was expanded in a total number of 6540 states. All parameters given
above were checked thoroughly in convergence tests.

\begin{figure}[t] 
    \begin{center} 
      \includegraphics[width=0.48\textwidth]{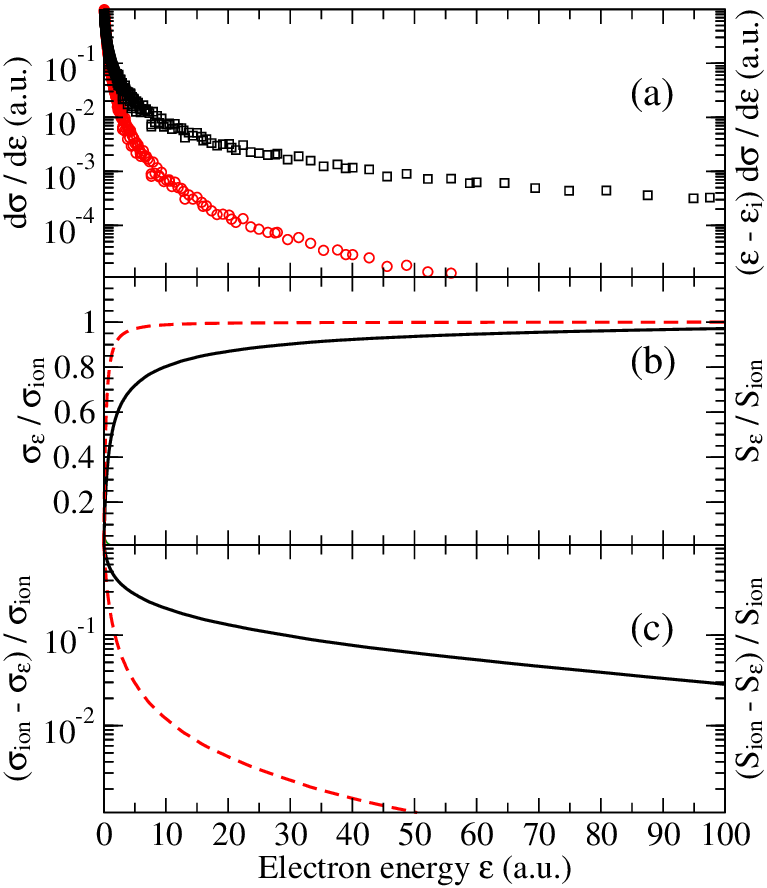} 
      \caption{(Color online) Convergence behavior of the ionization cross
        section and the contribution to the stopping power caused by
        ionization with respect to the 
        energy cutoff of the basis. The results are given for 3.2 MeV antiproton
        collisions with \hmol . 
        (a) 
        Electron energy spectrum d$\sigma$\,/\,d$\epsilon$, red circles;
        electron energy spectrum multiplied with the energy required for the
        excitation ($\epsilon-\epsilon_i$) d$\sigma$\,/\,d$\epsilon$ (see Eq.\
        (\ref{eq:stopping_power})), black squares.   
        (b) 
        Ratio $\sigma_\epsilon / \sigma_{\rm ion}$ of the ionization cross
        sections with an energy cutoff $\epsilon$ 
        to the one with the cutoff 250 a.u., red dashed line;
        ratio $S_\epsilon / S_{\rm ion}$  of the stopping power caused by
        ionization with an energy cutoff $\epsilon$ to the one with the 
        cutoff 250 a.u., black solid line.  
        (c) 
        The deviation from unity of the curves given in (b).
       \label{fig:Eloss_convergence} } 
    \end{center} 
\end{figure} 
\begin{table}[b]
  \centering
  \caption{Convergence with respect to the energy cutoff of the ionization cross
    section $\sigma_{\rm ion}$ and the contribution to the stopping power
    caused by ionization $S_{\rm ion}$ for 3.2 MeV antiproton 
    collisions with \hmol . Four different values for the energy cutoff are
    given which are sufficient to recover the final result (with an cutoff
    energy of 250 a.u.) within the given relative accuracy. }  
  \label{tab:Eloss_convergence}
  \begin{tabular}{c@{\hspace{0.75cm}}c@{\hspace{0.75cm}}c}
\hline
\hline
degree of recovery  & \multicolumn{2}{c}{energy cutoff $\epsilon$ }   \\
of final value      & $\sigma_{\rm ion}$ & $S_{\rm ion}$   \\
 ($\%$)         &   (a.u.) &    (a.u.)   \\ 
\hline
 90       &  1.8                &   29.0           \\ 
 95       &  3.2                &   63.5           \\ 
 97       &  4.8                &   96.5           \\  
 99       & 11.3                &   155            \\   
\hline
\hline  
  \end{tabular}
\end{table}

It was found that
especially a sufficiently high energy cutoff and density of continuum states
also at large state energies are of importance for converged results. 
This is somehow contrary to what is expected for ionization cross sections.
%
However, an insufficient choice of both parameters influences the final
stopping power differently which may even lead to some kind of compensation. A
too small energy cutoff results in a too small stopping power while an
improvement of the density of continuum states, on the other hand, led in the
present study to smaller stopping powers.

Figure 1 illustrates the convergence with respect to the
energy cutoff of the employed basis of the contribution to the stopping power
caused by ionization
$S_{\rm ion}$ (cf. Eq.\ (\ref{eq:total_ionization}))  and the ionization cross
section $\sigma_{\rm ion}$. The results 
are calculated for 3.2 MeV antiprotons colliding with \hmol . In Fig.\
\ref{fig:Eloss_convergence}(a) it can be seen that the electron energy
spectrum d$\sigma$\,/\,d$\epsilon$ decreases much faster for increasing
$\epsilon$ than the product $(\epsilon-\epsilon_i)
d\sigma/d\epsilon$. Therefore, the contribution to the stopping power caused by
ionization converges much slower with respect to the cutoff energy of the
basis than the ionization cross section, as can be seen in Fig.\
\ref{fig:Eloss_convergence}(b). Here, the quantities $\sigma_\epsilon$ and
$S_\epsilon$ only take transitions into final states $\phi_f$ with positive
$\epsilon_f \le \epsilon$ into account. Figure \ref{fig:Eloss_convergence}(c)
shows how much $\sigma_\epsilon$ and $S_\epsilon$ deviate from the final value
($\epsilon_f\le 250$ a.u.) when the cutoff energy is chosen as $\epsilon$. In
table \ref{tab:Eloss_convergence} those cutoff energies are given which
recover the final values of $\sigma_{\rm ion}$ and $S_{\rm ion}$ obtained with
a cutoff of 250 a.u.\ within 90, 95, 97, and 99\,\%.  
Figure \ref{fig:Eloss_convergence}(c) and table \ref{tab:Eloss_convergence}
clearly show the different convergence behavior of $\sigma_\epsilon$ and
$S_\epsilon$ with respect to the cutoff energy. While the ionization cross
section is converged within approximately 1\,\% with a cutoff of 10 a.u.\  in
the case of  $S_{\rm ion}$ a convergence within 3\,\% is only achieved with
an cutoff of around 100 a.u.

Obviously, this slow convergence behavior of $S_{\rm ion}$  becomes more
pronounced for higher impact energies since the relative population of
high-lying continuum states increases leading to a less steep fall-off of the
electron energy spectra as it was discussed in \cite{anti:luhr08a}. On the
other hand, for lower impact energies a smaller energy cutoff is sufficient
since the electron energy spectra fall off steeply for $\epsilon \ge
\frac{1}{2}(2 v)^2$ corresponding to the maximally transferred energy in a
classical collision 
\cite{anti:luhr08a}.  

%
\section{Results} 
\label{sec:results} 
%
%
%
%
%
%
%
%
%
%
%
%
%
%
%
%
Calculations were performed for \pb\ collisions with the three targets H,
\hmol , and He. The present data for the stopping power are listed in Table
\ref{tab:stopping_power}. In the following the findings of all three targets
will be separately discussed and compared with literature data.
%
%
%
\begin{table}[b]
  \centering
  \caption{Stopping power $S$ per atom for antiproton collisions with H, \hmol
    , and He in 10$^{-15}$ eV cm$^2$\,/\,atom which are shown in Figs.\
    \ref{fig:Eloss_H}, \ref{fig:Eloss_H2}, and \ref{fig:Eloss_He},
    respectively. The results for \hmol\ are given for the mean value of the
    internuclear distance $R_n=\mean{R_n}=1.4487$
    a.u.\ as proposed in \cite{anti:luhr08a}.}  
  \begin{tabular}{@{\hspace{0.1cm}}r@{\hspace{1.0cm}}c@{\hspace{0.5cm}}c@{\hspace{0.5cm}}c@{\hspace{0.1cm}}}
    \hline
    \hline
    $E$ (keV) & H      & \hmol     & He \\
    \hline
         1  &2.774        &2.242           &2.261    \\
         2  &3.164        &2.503           &2.524    \\
         4  &3.641        &2.826           &2.859    \\
         8  &4.208        &3.188           &3.280    \\
        16  &4.782        &3.596           &3.795    \\
        25  &5.098        &3.831           &4.144    \\
        32  &5.196        &3.947           &4.367    \\
        50  &5.210        &4.027           &4.666    \\
        64  &5.115        &4.013           &4.811    \\
       100  &4.623        &3.782           &4.850    \\
       128  &4.229        &3.557           &4.753    \\
       200  &3.406        &2.950           &4.316    \\
       256  &2.938        &2.588           &3.965    \\
       400  &2.148        &1.961           &3.202    \\
       800  &1.239        &1.165           &2.036    \\
      1600  &0.681        &0.656           &1.189    \\
      3200  &0.361        &0.356           &0.668    \\
      6400  &0.187        &0.187           &0.358    \\
    \hline
    \hline
  \end{tabular}
  \label{tab:stopping_power}
\end{table}
%

%
%
%
\subsection{$\bm{\bar{ p } }$ + H} 
\label{sec:pb_H} 
%
%

 %
\begin{figure}[t] 
    \begin{center} 
      \includegraphics[width=0.48\textwidth]{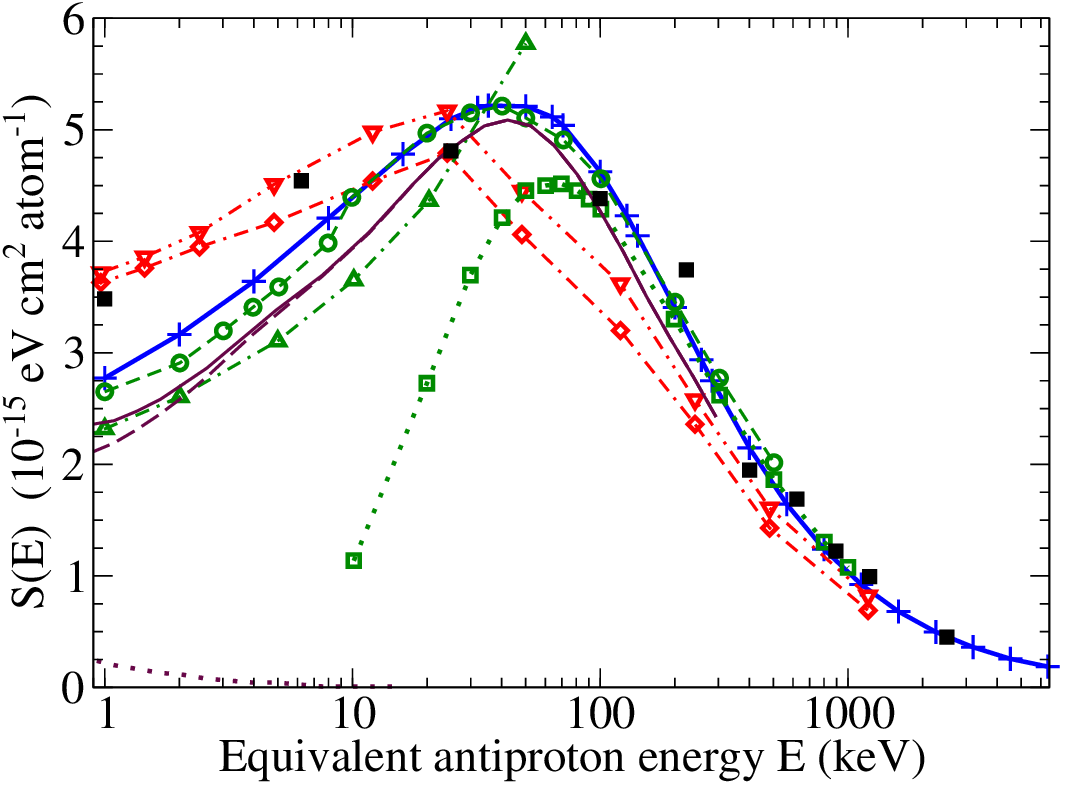} 
      \caption{(Color online) Energy-loss cross section $S(E)$ for H targets
       as a function of the equivalent antiproton impact energy $E$.   
       {\bf Theory}.
       Present results: blue solid curve with plus.
       Schiwietz {\it et al.}~\cite{anti:schi96}: 
       green dashed curve with circles, atomic orbital (AO);  
       green dash--dotted curve with triangles up, adiabatic-ionization (AI);
       green dotted curve with squares, distorted wave (DW).
       Cabrera-Trujillo {\it et al.}~\cite{anti:cabr05}:  
       brown thin solid curve, total $S$;  
       brown long dashed curve, electronic $S$;
       brown dotted curve, nuclear $S$.
       Custidiano and Jakas~\cite{anti:cust05}: black squares, CTMC for \pb .
       Cohen~\cite{anti:cohe83}: 
       red dash--doubly-dotted curve with triangles down, CTMC (CL) for
       $\mu^-$; 
       red doubly-dash--dotted curve with diamonds, quantum-classical CTMC
       (QC) for $\mu^-$.
       \label{fig:Eloss_H} } 
    \end{center} 
\end{figure} 

The stopping power for atomic hydrogen is shown in Fig.\
\ref{fig:Eloss_H}. Since no experiments have been performed for atomic 
hydrogen targets so far the present results are compared to various theoretical
calculations for \pb\ and \mm\ impact. The stopping power for hydrogen atoms is
preferably used for the testing of a theoretical approach since the target
description is well known and in principle no approximations are needed.  A
detailed analysis of $S$ for H and He was  done by Schiwietz {\it et al.}\
\cite{anti:schi96} comparing three different approaches, namely, an
atomic-orbital (AO), a distorted-wave (DW), and an adiabatic-ionization (AI)
description. Due to the inherent approximations of the AI --- adiabatic
collision --- and the DW --- interaction in first order  --- approaches they are
basically low-energy and high-energy methods,  respectively. Their advantage
over the AO method is based on their comparably small numerical effort. The AI
and DW results describe the stopping power reasonably for $E<20$ keV and
$E>100$ keV, respectively. Note, that the use of the DW method
leads to a clearly different position and height of the stopping maximum
compared to the AO method and the AI curve does not show any maximum at all. 
The present findings, which are also based on an atomic-orbital approach, are
in good agreement with the AO results, except 
for the regime 2 keV $<E<8$ keV where a small discrepancy exists. From the
comparison to the AO results it is assumed that the present method is
correctly implemented.   

Cabrera-Trujillo {\it et al.}\ \cite{anti:cabr05}
employed the electron-nuclear dynamics (END) theory which is based on the
application of the time-dependent variational principle to the Schr\"odinger
equation using a coherent state representation of the wave function. This
method allows for the simultaneous determination of the electronic and
nuclear stopping power. The latter is small for all surveyed projectile
energies and completely negligible for $E>10$ keV. The END results for the
electronic stopping power show a similar behavior like both AO calculations
but predict throughout lower values. These three curves share in particular the
position of the maximum at around $\emax\approx 40$ keV and similar slopes for
energies below and above $\emax$. 

The Classical Trajectory Monte Carlo (CTMC) method was recently employed by
Custidiano and Jakas \cite{anti:cust05} in order to determine  \spb\ and earlier
already by Cohen \cite{anti:cohe83} for \smm. Both calculations agree
for high energies $E>200$ keV with the AO, DW, and END 
results but differ from them below the stopping power maximum sharing the same
slope. While the CTMC results for \spb\ follow the trend of the AO and END
curves down to about 20 keV the \smm\ results by Cohen show a different
behavior in the energy range around $\emax$. Besides the
purely classical CTMC (CL) Cohen also provided a quantum-classical analysis
(QC) of his data. They differ mainly in the vicinity of $\emax$ where the
CL results are closer to the END and AO curves than those from the QC
analysis. It was shown in \cite{anti:cust05} that for low impact energies
$E<30$ keV the CTMC stopping power depends considerably on the eccentricity of
the initial classical electron orbits. The similar behavior of all CTMC results
below 30 keV may be caused by the fact that Custidiano and Jakas followed a
procedure for preparing initial conditions described by Cohen. 

Finally, it is possible to conclude that the present findings for H targets
agree well 
with the other AO calculation and share the same behavior than the END
results. For all other approaches considered here the energy range in which they
are applicable is limited to energies around and above the stopping maximum
$E\gtrapprox\emax$ 
except for the AI method which gives reasonable results only below the maximum.
%
%
%
\subsection{$\bm{\bar{ p } }$ + \hmol } 
\label{sec:pb_H2} 
%
%
%
%
%
%
%
%
%
%
 %
\begin{figure}[t] 
    \begin{center} 
      \includegraphics[width=0.48\textwidth]{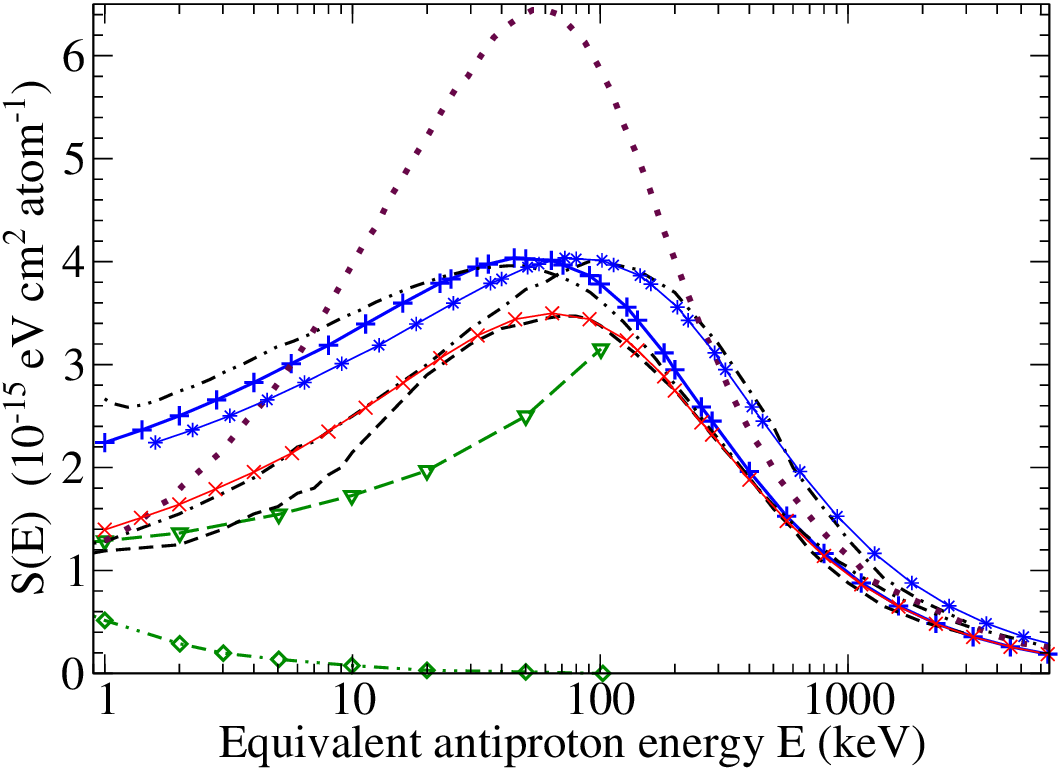} 
      \caption{(Color online)  Energy-loss cross section $S(E)$ for \hmol\
       targets as a function of the equivalent antiproton impact energy $E$.   
       {\bf Theory}.
       Present results: blue solid curve with plus, $R_n$=1.4487;
       blue thin solid curve with stars, $E$ scaled by a factor 1.6 (see text);
       red thin solid curve with x, double ionization excluded (see text).
       Schiwietz {\it et al.}~\cite{anti:schi96}: 
       green dashed curve with triangles down, adiabatic-ionization (AI);  
       green dash--doubly-dotted curve with diamonds, nuclear stopping.
       {\bf Experiment}. \pb :
       Agnello {\it et al.}~\cite{anti:agne95}, black dash--dotted curve;
       Adamo {\it et al.}~\cite{anti:adam93},   black dash--doubly-dotted curve.
       \mm : Hauser {\it et al.}~\cite{sct:haus93}, black dashed curve.
       $p$: Andersen and Ziegler \cite{sct:ande77}, brown dotted curve.
       \label{fig:Eloss_H2} } 
    \end{center} 
\end{figure} 
In Fig.\ \ref{fig:Eloss_H2} the $S(E)$ for \hmol\ targets is shown as a
function of the equivalent antiproton impact energy. The equivalent antiproton
energy can be obtained by multiplication of the impact energy with the factor
$m_{\bar{p}}/M_p$ where $m_{\bar{p}}$ is the mass of an antiproton. The factor
for \mm{}\ projectiles is accordingly $m_{\bar{p}}/m_{\mu\, \bar{}} \approx
8.880$. The present data are  
calculated for a fixed internuclear distance $R_n=\mean{R_n}=1.4487$ a.u.\ of
the two nuclei as proposed in \cite{anti:luhr08a,nu:saen97b}.
The consideration of isotopes
of hydrogen molecules leads to a slightly different $\mean{R_n}$
\cite{dia:kolo64}. The effect on the stopping power for different $R_n$ in the
range 1.4 a.u. $\le R_n\le$ 1.5 a.u.\ is only quantitative and largest around
the stopping maximum where the deviation is of the order of $1.5\,\%$.   

In contrast to the 
hydrogen atom three experiments have been performed for \hmol\ whereas the
authors are only aware of a single calculation in the molecular case by
Schiwietz {\it et al.}\ \cite{anti:schi96} employing the AI method in which
the \hmol\ molecule was described in a quasiatomic way with a single effective
scaled charge. Within an IPM the effective charge was
chosen in such a way that the total electronic energy of the molecule at its
equilibrium internuclear distance is reproduced.

It is evident form Fig.\ \ref{fig:Eloss_H2} that the experimental results for
\spb\ by Adamo {\it et al.}\ \cite{anti:adam93} and by Agnello {\it et al.}\
\cite{anti:agne95} as well as for  \smm\ by Hauser {\it et al.}\
\cite{sct:haus93} all differ considerably. For high energies $E>200$ keV the
findings for \pb\ impact by Adamo {\it et al.}\ and \mm\ impact are very
similar. At energies below the maximum the \smm\ are closer to the
more recent 
\pb\ results by Agnello {\it et al.} The maxima of the three experimental
curves  \cite{anti:adam93}, \cite{anti:agne95}, and \cite{sct:haus93} are
situated approximately at the equivalent antiproton energies $\emax\approx 45$
keV, 100 keV, and 75 keV, respectively. While the maxima of both \spb\ curves
are of comparable height the maximum of \smm\ lies well below those two. 

It should be noted that the experimental curves shown here are the best fit
results from an analysis of the measured data. The order of the uncertainties
was estimated in \cite{anti:agne95} to amount to $\pm$10\%.  In the case of
the \mm\ results
the uncertainties vary from $\pm$10\% for impact energies in the vicinity and
above $\emax$ and increase up to $\pm$50\% for decreasing $E$.
Furthermore, in the \mm\ experiments basically the excitation cross section
was determined only as stated already in the introduction. The shown \smm\
results depend therefore also on additional 
data which were taken from literature. The experimental \pb\ ionization
cross sections  $\sigma_{\rm ion}$  \cite{anti:ande90a} used in order to
determine \smm , however, were later on found to be erroneous for $E<200$ keV
\cite{anti:hvel94,anti:knud08a}.  

Due to these substantial uncertainties it is one aim of this work to
discriminate with the help of the present findings between the different
experimental results. For $E>200$ keV  the present results are in good
agreement with the  $\mu^-$ data and the \spb\ by Adamo {\it et al.}\ While the
latter curve has a similar behavior like the present calculations also in the
vicinity and below the maximum the former \smm\ curve deviates clearly for
$E<200$ keV. The \spb\ curve determined by Agnello {\it et al.}\ is on the other
hand not compatible with the present data. While the height of both maxima is
very similar it appears as if the experimental curve is shifted to larger
energies.  A simple scaling of $E$ by a factor of 1.6 between the present
and the experimental \pb\ data by Agnello {\it et al.}\ as proposed in
\cite{anti:luhr09b} can, however, not be verified in view of the current more
detailed investigation. The scaled curve clearly deviates from the measured
data for the highest energies. On the other hand, it is the scaled curve which
looks most similar to the one for \spb\ by Agnello {\it et al.}\ for impact
energies around and above the stopping maximum.

It is known that the IPM overestimates the
two-electron processes like double ionization (e.g., \cite{anti:wehr96}) which
was also observed in an earlier application of the model potential
\cite{dia:luhr08}. Single excitation and single ionization are on the other
hand reasonably well described. Therefore, the present stopping power without
the contribution from double ionization \ssin\ has also been analyzed 
by using the difference between Eq.\ (\ref{eq:de_final}) and Eq.\
(\ref{eq:de_ionization_final}) instead of the total $S$ given by Eq.\ 
(\ref{eq:de_final}).   
The qualitative behavior of the present curves for
$S$ and \ssin\ is similar due to the fact that both curves originate from the
same calculation. The quantitative difference on the other hand increases 
for low impact energies. While the relative difference is below 1\,\% for
$E>1500$ keV it is larger than 10\,\% for $E<100$ keV and finally becomes as
large as one third for $E=2$ keV.
In the validity range of the used model this curve can be interpreted as a
lower bound to the stopping power. For $E>40$ keV \ssin\ matches the
experimental \smm\ while for $E<25$ keV the experimental data by Agnello {\it
  et al.}\ are reproduced by \ssin . 
Unfortunately, the authors are not aware
of any independent and reliable data for the single and especially double
ionization or excitation cross section for low-energy \pb\ + \hmol\
collisions. These would allow for a quantitative approximation of the
uncertainties due to the model potential and the use of the IPM for impact
energies below 100 keV.

The other theoretical curve calculated by Schiwietz {\it et al.}\ shows a
similar dependence on $E$ like the AI results in the case for atomic H
targets. It agrees with the measurements of \spb\ by Agnello {\it et al.}\
and of \smm\  by Hauser  {\it et al.}\ for $E<5$ keV but differs clearly for
$E>10$ keV from all other curves. The nuclear stopping power also calculated
by Schiwietz {\it et al.}\ \cite{anti:schi96} is again small in the considered
energy regime but considerably larger than the END results in the case of
atomic H.

A comparison to the stopping power for $p$ impact shows a maximum at $E\approx
60$ keV which is about 60\% larger than the present value for \spb .  
Note that all curves in Fig.\ \ref{fig:Eloss_H2} lie below the
$p$ results for energies larger than $\emax$ except for \spb\
determined by Agnello {\it et al.} At high energies all curves converge to the
$p$ results showing a $1/v^2$ dependence of $S$ as expected form the Bethe
theory (cf. Eq.\ (\ref{eq:stopping})). Below the
maximum  only the present curve and the one by Adamo {\it et al.}\ cross the
$p$ curve for $E>1$ keV resulting in a change of the sign of the Barkas term. 

%
%
%
\subsection{$\bm{\bar{ p } }$ + He}  
\label{sec:pb_He} 
%
%
%
\begin{figure}[t] 
    \begin{center} 
      \includegraphics[width=0.48\textwidth]{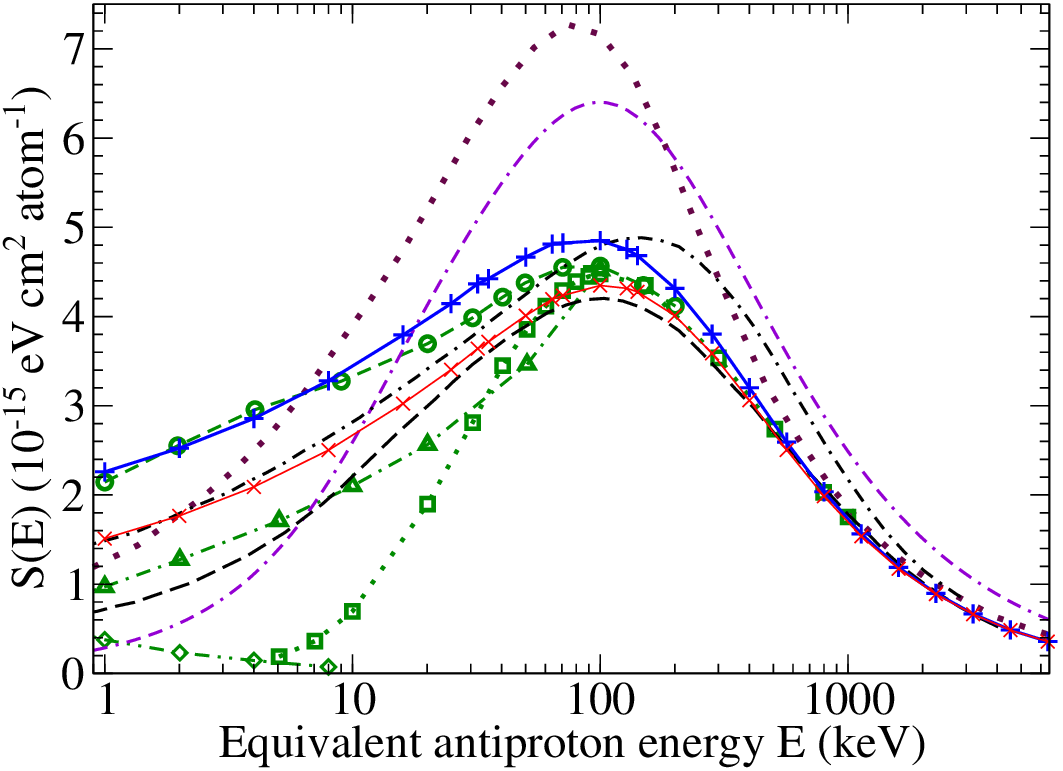} 
      \caption{(Color online) Energy-loss cross section $S(E)$ for He targets
       as a function of the equivalent antiproton impact energy $E$.    
       {\bf Theory}.
       Present results: blue solid curve with plus;
       red thin solid curve with x, double ionization excluded (see text).
       Schiwietz {\it et al.}~\cite{anti:schi96}: 
       green dashed curve with circles, atomic orbital (AO);  
       green dash--dotted curve with triangles up, adiabatic-ionization (AI);
       green dotted curve with squares, distorted wave (DW);
       green dash--doubly-dotted curve with diamonds, nuclear stopping.
       Basko \cite{sct:bask05}: violet doubly-dash--dotted curve,
       low-velocity Bohr (LVB).
       {\bf Experiment}.
       \pb\,: Agnello {\it et al.}~\cite{anti:agne95}, black dash--dotted curve.
       \mm : Kottmann~\cite{sct:kott87}: black dashed curve.
       $p$: Andersen and Ziegler \cite{sct:ande77}, brown dotted curve.
       \label{fig:Eloss_He} }    
    \end{center} 
\end{figure} 
In contrast to hydrogen, data of more than one experimental and theoretical
approach exist for He targets. They are shown in Fig.~\ref{fig:Eloss_He}
together with the present findings and the experimental results for $p$
impact. For \pb\ + He collisions also the ionization cross section
$\sigma_{\rm ion}$ is experimentally and especially theoretically well studied
(cf. \cite{anti:knud08} and references therein) making it a good candidate for
the comparison of different approaches.

The experimental curves for \pb\ and \mm\ stopping in He gases show a
behavior similar to the one of the \hmol\ target measured by the same groups
which were shown in Fig.~\ref{fig:Eloss_H2}. The stopping maximum is 
approximately 25\% higher for He than for \hmol . Again the experimental \spb\
by Agnello \cite{anti:agne95} {\it et al.}\ is larger than the $p$ results by
Andersen and Ziegler \cite{sct:ande77} above the stopping maximum while the
measured \smm\ by Kottmann \cite{sct:kott87} stays below the $p$ curve for all
energies considered here.  

As for the atomic H target Schiwietz {\it et al.}\ \cite{anti:schi96} applied
the AI, AO, and DW method to calculate the He stopping power. The AI curve
shows a functional dependence on $E$ analogous to the one observed for H and
\hmol\ targets. That is, for small $E$ it is generally in accordance with the
two experimental \spb\ and \smm\ curves while it seems not to be applicable for
$E>20$ keV. The DW results
fully agree with the experimental \smm\ for $E>300$ keV but fall off much
faster below the stopping maximum for $E<40$ keV. Exactly the same behavior was
observed for $\sigma_{\rm ion}$ calculated earlier by Fainstein {\it et
  al.}~\cite{anti:fain87} also using a DW method which coincides with the
$\sigma_{\rm ion}$ resulting from the DW calculations by Schiwietz {\it et
  al.}\ \cite{anti:schi96}. Although first 
measurements of low-energy ionization for \pb\ + He collisions
\cite{anti:ande90,anti:hvel94} fully confirmed this steep fall off below the
ionization maximum, a recent more accurate experiment was able to clearly
contradict this trend \cite{anti:knud08} in favor of a less steep decrease of
$\sigma_{\rm ion}$ 
below the maximum. The $S$ results calculated by Schiwietz {\it et al.}\ using
the AO and the DW method fully agree with each other for $100<E<200$ keV both
having a maximum value lying in between the two experimental curves at
$E\approx100$ 
keV. This is somehow different from the case of atomic H targets where the
height and position of the AO and DW stopping maxima clearly differ (see Fig.\
\ref{fig:Eloss_H}). Below the
maximum, however, the two curves diverge with decreasing  
$E$. The AO results stay above the DW and the experimental data with
deviations increasing to more than 50\% for $E<5$ keV. These deviations of the
AO results were explained by the use of a model treating one active electron
in the effective potential of the heavy nucleus and a static density
distribution of the second inactive electron which screens the nucleus
\cite{anti:schi96}. 
In the adiabatic limit of the AO model for He no ionization
threshold exists for $R\rightarrow 0$ as it is known for an H atom also
referred to as Fermi-Teller radius. This is, however, in contrast to a full
two-electron treatment of a He atom which leads for $R\rightarrow 0$ to a
finite ionization threshold of $\approx$\,0.7 eV due to the fact that the
electron density is changed dynamically when the \pb\ approaches the nucleus.
Therefore, the AO results were expected to overestimate the
ionization cross section and consequently also the stopping power for low $E$
\cite{anti:schi96}.

The present results for $S$ coincide with the experimental \smm\ and
theoretical DW data for high energies $E>500$ keV but become considerably
larger for $E<200$ keV. Like for the \hmol\ target  the maximum of the present
He curve has the same height as the \spb\ measured by Agnello {\it et al.}\
and it is situated  around 100 keV as predicted by the AO and DW
methods. Below the stopping 
maximum the present data are, however, much larger than the experimental \spb\
and \smm\  and theoretical AI results. The present calculations are, on the
other hand, basically in agreement with the AO data for low
energies. Therefore, it may be concluded that the deviations of the present
findings at low energies also originate from deficiencies of the employed
effective one-electron model which lead to substantial changes of the
ionization potential in the adiabatic limit as is the case for the
AO model. Due to the existing uncertainties of experimental and theoretical
results, especially for low energies, it is, however, not possible to finally
conclude on the exact behavior of the stopping power in this energy range. On
the other  hand, the error of the measured \smm\ curve could be reduced
drastically if a point at low energies could be fixed safely
\cite{sct:kott87}. In this context it would be valuable to perform a
calculation using a full two-electron description of the target to eliminate
the uncertainties connected so far with both AO approaches using effective
one-electron models. 
 
As has been done for \hmol\ the stopping power excluding double ionization
\ssin\ has also been analyzed for He targets. 
The present  $S$ and \ssin\ curves are again qualitatively similar while
quantitative differences increase for lower impact energies.
The relative contribution from double ionization to $S$ is at low energies
slightly smaller than in the case of \hmol\ while it is the other way round at
high energies.  
It is interesting to note that these relative
contributions correspond roughly to the ratios of cross sections for double
and single ionization in the IPM (e.g., \cite{anti:wehr96}) multiplied by
two. The factor two accounts for the energy of both electrons involved in the
double ionization. 
The \ssin\ curve lies above the experimental data for \mm\ {}  for $E<400$ keV
indicating that the measured results might be too small. For $E<25$ keV \ssin\
describes the experimental \pb\ data reasonable as it is the case for \hmol\
pointing out that the contribution of double ionization in $S$ is too large
especially at low energies.

In contrast to the case of \pb\ + \hmol\ a number of advanced calculations
(e.g., \cite{anti:fost08,anti:igar04,anti:read97})
were performed for \pb\ + He ionization cross sections in addition to the
experiments. This allows for a rough estimate of some of the uncertainties of
the present results stemming from the target model and the use of the IPM for
the presented He stopping power as well as for an attempt to estimate an
corrected value of $S$ at low impact energies. The
present cross section for single ionization is in good agreement with
experiment for $E>40$ keV but becomes increasingly too large for smaller
impact energies due to the reasons discussed above. The cross section for
double ionization depends quadratically on the single-electron
ionization probability within the IPM which is, however, known to overestimate
the measured data (cf., e.g., \cite{anti:wehr96}). In the following, the
averaged energy transfer is assumed to be described correctly.  Then the
stopping power depends linearly on the cross sections. Under this assumption
the correct contribution to the stopping power due to double ionization may be
roughly approximated as being only  50\%, 43\%, and 30\% of the difference
$S-S_{\rm sin}$ for the three energies 200, 100, and 25 keV, respectively. The
single ionization cross section of a recent accurate calculation
\cite{anti:fost08} is approximately 10\% smaller than the present one for
$E=25$ keV. For this energy a value of the present stopping power which
includes all mentioned assumptions and corrections may be therefore 
roughly approximated with 3.3 $10^{-15}$ eV cm$^2$ per atom. This value lies
slightly below the curve for $S_{\rm sin}$. 
Although no quantitative estimate can be done for \hmol\ targets as discussed
above it might be expected that the correction is qualitatively similar to that
performed for He. 
Note that the contributions due to excitation have not been changed in this
simple estimate. The above discussion obviously shows the need for further
calculations using a two-electron description of the target in order to
improve the quantitative description at impact energies below 100 keV.  
%

Also shown in Fig.\ \ref{fig:Eloss_He} is a calculation by Basko using a
semi-classical low-velocity Bohr (LVB) stopping model \cite{sct:bask05}
that extends the Bohr model to lower energies in which the stopping number $L$
depends on the sign of the projectile.  
The shape of the LVB curve is similar to that for $p$ but shifted to higher
energies. 
Besides the \spb\ for \hmol\ and He measured by Agnello {\it et al.} the LVB
curve is the only one with values larger than for
$p$ impact for energies above the maximum. Below the stopping maximum the LVB
results stay well below the $p$ results and cross all other curves. Besides
the position of the stopping maximum the outcome of the LVB method does not
fit well any of the curves discussed here. 

%
%
%
\section{Discussion}
\label{sec:discussion} 
%
%
%
%

%
%
%
%
 \subsection{The Barkas effect} 

From Eqs.\ (\ref{eq:stopping}) and (\ref{eq:born_series}) as well as the
discussed experimental evidence it is apparent that higher-order terms in
$Z_p$ will 
be present in an exact calculation of the stopping power. In order to
highlight the Barkas effect and also higher orders of $L$ in $S$ it is common
to determine the relative stopping power for particles and their antiparticles
\begin{equation}
  \label{eq:DeltaS_S}
  \frac{\Delta S}{S} = \frac{S^p - S^{\bar{p}}}{S^p}\,.
\end{equation}
Using Eqs.\ (\ref{eq:stopping}) and (\ref{eq:born_series}) Eq.\
(\ref{eq:DeltaS_S}) can be rewritten as
\begin{equation}
  \label{eq:DeltaS_S_L1}
  \frac{\Delta S}{S} =
  \frac{2Z_pL_1+2(Z_p)^3L_3+\ldots}{L_0+Z_pL_1+(Z_p)^2L_2+\ldots} 
\end{equation}
showing that it depends only on odd terms. In the case that
higher-order terms are insignificant (i.e., $|L_{2i+1}|\ll |L_1|$ , $i>0$)
${\Delta S}\,/\,{S}$ becomes approximately proportional to the Barkas term
$L_B$. Then the first-order correction $L_B$ can be approximated, using
Eqs.\ (\ref{eq:stopping}) and (\ref{eq:DeltaS_S_L1}), by
\begin{equation}
  \label{eq:L1}
  L_B \approx \sum_{i=0}^{\infty} L_{2i+1}(Z_p)^{2i}=\frac{1}{8\pi (Z_p)^3 Z}\,v^2\Delta S\,.
\end{equation}
Strictly speaking, $L_B$ is equal to the correction to the stopping number due
to the sum of all odd terms $L_{2i+1}$ since the projectile charges considered
in this work have the absolute value $|Z_p|=1$. Therefore, $L_B$ can be
considered soundly also, if the condition that $L_1$ is the
dominant odd contribution is not fulfilled. Though, in that case it is not
appropriate to call $L_B$  Barkas term.

\begin{figure}[t] 
    \begin{center} 
      \includegraphics[width=0.48\textwidth]{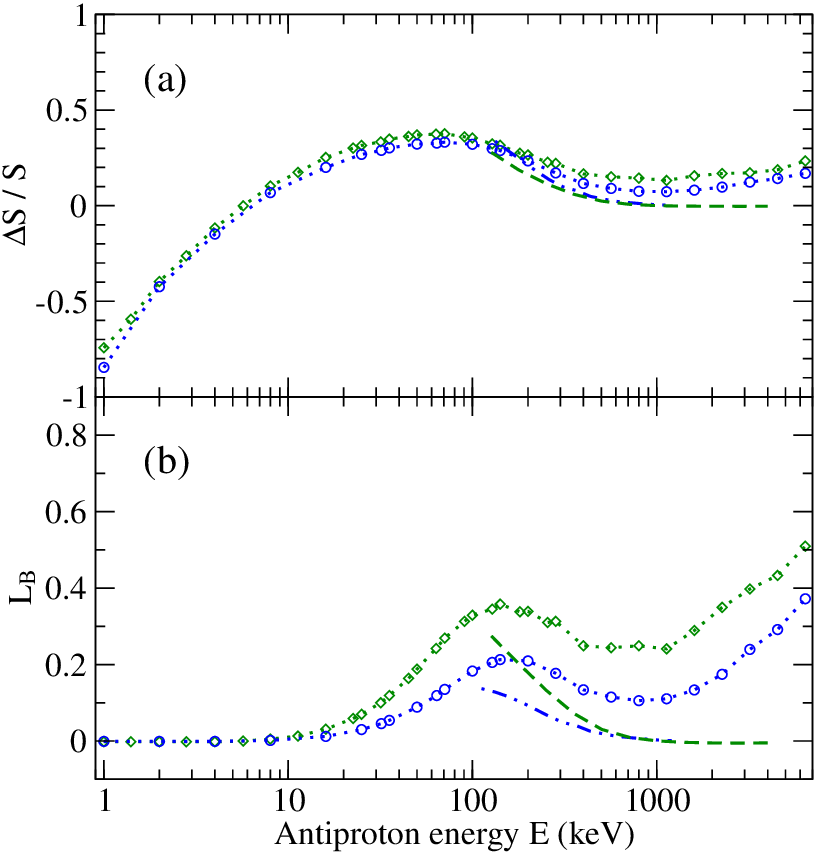} 
      \caption{(Color online) (a) Relative stopping power $\Delta
        S\,/\,S$ for antiprotons and protons colliding with \hmol\ and He
        as a function of the impact energy $E$. (b) First order correction to
        the stopping number $L(E)$ also referred to as Barkas correction
        $L_B(E)$ as given in Eq.\ (\ref{eq:L1}) (see text). 
        Comparison of present \spb\ with experimental $S^p$ measured by
        Andersen and Ziegler \cite{sct:ande77}:
        green diamonds, \hmol\ molecule;
        blue circles, He atom.
        Comparison of present \spb\ and $S^p$:
        green dashed curve, \hmol\ molecule;
        blue dash--doubly-dotted curve, He atom.
       \label{fig:DeltaS_S} }    
    \end{center} 
\end{figure} 

The present results for  ${\Delta  S}\,/\,{S}$ are shown in Fig.\
\ref{fig:DeltaS_S}(a) for \hmol\ and He. In order to determine the stopping
ratios the calculations for antiprotons were compared to the experimental data
for proton collisions \cite{sct:ande77}. It can be seen that the ratios for
\hmol\ and He show a comparable behavior. The ratio increases from
about $-0.8$ at 1 keV to a maximal value of approximately 0.35 at 70 keV and
than starts to fall off. However, for $E>1000$ keV  ${\Delta  S}\,/\,{S}$
begins to increase again for increasing $E$. The ${\Delta
  S}\,/\,{S}$ curve for an H atom calculated by Cabrera-Trujillo {\it et al.}\
for $E\le 300$ keV \cite{anti:cabr05} shows the same qualitative behavior. The
outcome from the calculations by Schiwietz {\it el al.} suggests a decreasing
ratio from the stopping maximum until their highest calculated impact energy
$E=1000$ keV.  
An increase of the ratio for high energies is not expected since in the limit
of high impact energies the first Born approximation is known to give
satisfying results.  

The calculated \spb\ for \hmol\  and He are also compared
to present results for $S^p$ for high impact energies in Fig.\
\ref{fig:DeltaS_S}(a). For $E>200$ keV the electron capture cross section
becomes negligible \cite{sct:shah85,sct:shin89}. 
Therefore, the employed one-center approach is  
also capable to describe the stopping of $p$ for high energies. The relative
stopping power  ${\Delta  S}\,/\,{S}$ using only the present data for \pb\ and
$p$ impact decreases for increasing $E$. This means that the obtained \spb\
results are consistent within the employed model but deviate slightly from the
experimental $S^p$ data measured by Andersen and Ziegler \cite{sct:ande77}.
In the case that the used experimental proton data are
taken as reference that may indicate that the present ratios ${\Delta
  S}\,/\,{S}$  are not sufficiently converged
at high energies although the stopping power $S$ itself compares satisfactory
with the experimental results. On the other hand, the present results for
\spb\ did not change even with an increase of the energy cutoff by a factor of
20. An even further enlargement of the energy cutoff of the chosen basis set
would, however, lead to a drastic increase of the computational time due to the
fast oscillating phases in the differential equations for the expansion
coefficients $c_j({\bf R}(t))$ and was therefore not performed.


The present findings for  $L_B$ as given in Eq.\ (\ref{eq:L1}) are shown in
Fig.\ \ref{fig:DeltaS_S}(b) for \hmol\ and He. The qualitative behavior is
similar as for the ratio  ${\Delta  S}\,/\,{S}$ but they differ in the scaling
for different $E$. Since $L_B$ is proportional to $(v)^2$ and therefore to
$E$ it is suppressed at low $E$ but enhanced at high $E$.
The difference between \hmol\ and He is of the order of a factor two. Exactly
this factor enters in Eq.\ (\ref{eq:L1}) as the atomic number $Z$ in the
nominator being one for hydrogen and two for He. For the highest energies
$E>1000$ keV  $L_B$ determined with the experimental $S^p$ data increases also
with $E$ due to its proportionality to ${\Delta  S}\,/\,{S}$. On the other
hand, the $L_B$ curves evaluated only from the present results for \spb\ and
$S^p$ decrease with increasing $E$ and, as expected, approach zero for
$E>1000$ keV.
%
%
%
For small energies $E<10$ keV  $L_B$ is very small but non
zero except for $E\approx 6$ keV where it changes the sign. However, for these
small energies a sizable difference between the stopping power of \pb\ and 
$p$ exists.
This means  that the often used condition ($L_{2i+1} \ll 1$ for $i \ge 0$) to
assume that higher-order terms of the stopping number, which lead to different
results for particles and antiparticles, are insignificant is not
sufficient. While the sum of all odd corrections $L_B$ already fulfills
this condition in Fig.\ \ref{fig:DeltaS_S}(b) for small $E$ the findings for
the stopping power of $p$ and \pb\ impact clearly differ in Figs.\
\ref{fig:Eloss_H2}  and \ref{fig:Eloss_He}.

%
%
%
%
%
%
%
%

%
%
%
%
%
%
%
%
\subsection{Excitation energies and ratio  $\bm{S_{\rm exc} / S_{\rm ion}$}} 
\begin{figure}[t] 
    \begin{center} 
      \includegraphics[width=0.48\textwidth]{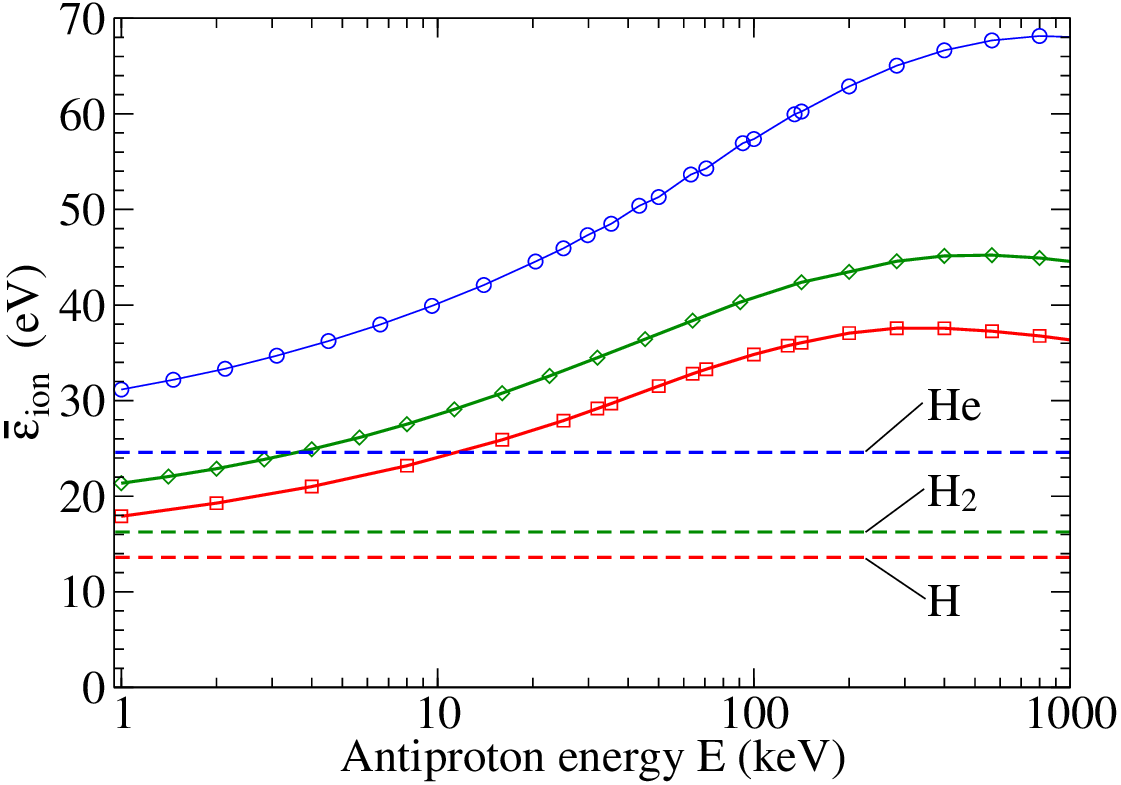} 
      \caption{(Color online) The average energy transfer to the ionized
        electrons \aei{} per electron as a function of the \pb\ impact energy.
        Red squares, H atom;
        green diamonds, \hmol\ molecule;
        blue circles, He atom.
        The ionization potential --- being the lower bound --- for the three
        targets are shown as dashed horizontal lines. 
       \label{fig:Eion} }    
    \end{center} 
\end{figure} 
\begin{figure}[t] 
    \begin{center} 
      \includegraphics[width=0.48\textwidth]{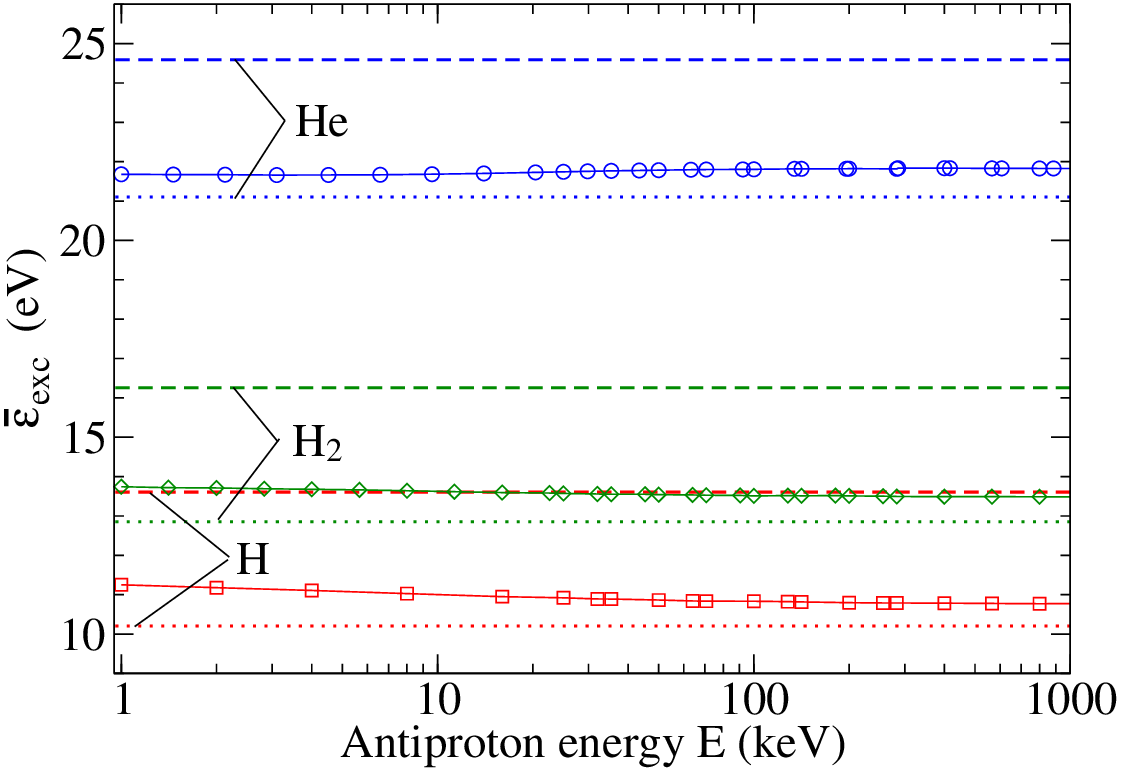} 
      \caption{(Color online) The average energy transfer to the bound
        electrons \aee{} per electron as a function of the \pb\ impact energy.
        Red squares, H atom;
        green diamonds, \hmol\ molecule;
        blue circles, He atom.
        The ionization potentials --- upper bound --- are shown as
        dashed horizontal lines and the excitation energy into the lowest
        dipole-allowed states --- lower bound --- as dotted horizontal lines. 
       \label{fig:Eex} }    
    \end{center} 
\end{figure} 
\begin{figure}[t] 
    \begin{center} 
      \includegraphics[width=0.48\textwidth]{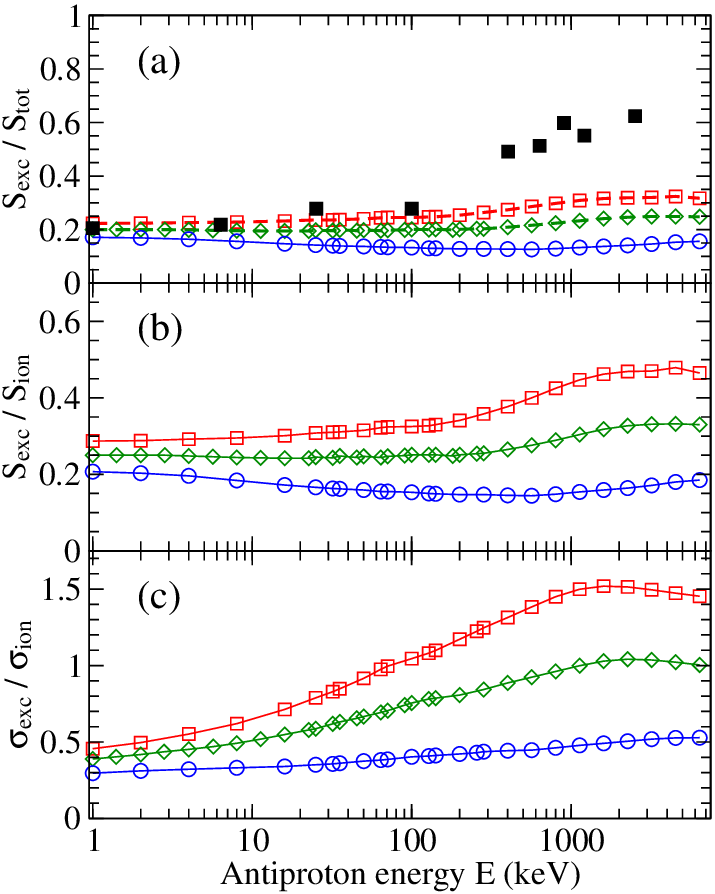} 
      \caption{(Color online) Ratios of the stopping power as a function of
        the \pb\ impact energy. (a) Ratio of stopping power due to excitation
        $S_{\rm exc}$ and total stopping power $S_{\rm tot} = S$. (b) Ratio of
        $S_{\rm exc}$ and stopping due to ionization $S_{\rm ion}$. (c) Ratio
        of cross sections for excitation $\sigma_{\rm exc}$ and ionization
        $\sigma_{\rm ion}$. 
        Present data:
        red squares, H atom;
        green diamonds, \hmol\ molecule;
        blue circles, He atom.
        Custidiano and Jakas \cite{anti:cust05}: 
        black filled squares, H atom with  CTMC. 
        \label{fig:Sexc_Sion} }    
    \end{center} 
\end{figure} 

For the correct determination and understanding of the stopping power the
different energy-loss processes are of interest.
Besides the cross sections for ionization and excitation which are discussed
elsewhere (cf.\ the references in Sec.\ \ref{sec:introduction}) also the
energy transfer is of importance. Thereby, quantities 
like the electron-energy spectra \cite{anti:luhr08a} or differential
excitation cross sections for transitions into single states \cite{dia:luhr08}
provide detailed insight. It is, however, also conclusive to look at the average
energies transferred to ionized and bound-excited target atoms. Knowing these
quantities it is possible to determine the stopping power out of total cross
sections for ionization and excitation. 

The average energy transfer to the ionized target atoms or molecules \aei{}
per electron as a function of 
the impact energy of the projectile $E$ for the three targets H, \hmol , and
He are shown in Fig.\ \ref{fig:Eion}. Also given are the
ionization potentials of the three targets being the lower limit of the energy
transfer which is required for ionization. The ionization potentials $I$ are
ordered as $I^{\rm He}>I^{{\rm H}_2}>I^{\rm H}$. The energy transfer to ionized
electrons is ordered in the same way \aei{\,{\rm He}}\  $>$ \aei{\,{\rm H}_2}\
$>$ \aei{\,{\rm H}}. All \ae{} increase for increasing impact energy $E$. This
trend is in agreement with the previous analysis of the electron-energy
spectra for \hmol\ \cite{anti:luhr08a,anti:luhr09b}. The increasing importance
of the higher-lying unbound states requires on the other hand also a
sufficient description of the continuum states at large electron energies
$\epsilon$. It turned out that the results for large $E$ become sensitive to
the energy cutoff of the employed basis set. Therefore, it is important at high
impact energies $E$ to extend the continuum part of the basis with continuum
states belonging to higher electron energies $\epsilon$ in order to achieve 
convergence. 

The average energy transfer to the excited atoms or molecules \aee{}  shown in
Fig.\ \ref{fig:Eex} is on the other hand only weakly dependent on the impact
energy $E$. The \aee{} curves for the three targets are energetically ordered
in the same way as the \aei{}\,.  Also given in Fig.\ \ref{fig:Eex} are the
ionization potentials being the upper limits for bound state
transitions as well as the 
minimum energy transfer into the first excited states for the
three targets which are all independent of $E$. The \aee{} curves for all
three targets stay close to the minimum lines for all $E$.
This is in accordance to the fact that the first excited dipole-allowed state
is the dominant excitation channel as was observed in
\cite{anti:luhr08,anti:luhr08a,anti:luhr09b} and is in agreement with
measurements for $e\, \bar{}$ + \hmol\ collisions \cite{sct:liu98}.  For
decreasing $E$ the contribution of the higher excited states increases for H
and \hmol\ while it decreases slightly in the case of He.

In Fig.\ \ref{fig:Sexc_Sion} the relative contribution to the stopping power
due to ionization and excitation is considered according to the Eqs.\
(\ref{eq:total_ionization}) and (\ref{eq:total_excitation}). The relative
importance of 
both processes excitation and ionization, depends on the target and the
impact energy of the \pb .  In general it can be concluded
that in the whole energy range the energy loss due to ionization dominates the
loss due to excitation as can be seen in Fig.\ \ref{fig:Sexc_Sion}(b). This
is in contrast to the findings for alkali-metal atoms where the \pb\ loses
energy mainly due to the bound-state excitation \cite{anti:luhr09a}.  

The fraction of
energy which goes into excitation given in Fig.\  \ref{fig:Sexc_Sion}(a) is
largest for H  and smallest for He in accordance with the corresponding ratios
of excitation to ionization cross section shown in \ref{fig:Sexc_Sion}(c). This
may be linked to the ionization potential which is smallest for H and largest
for He.

The present ratio $S_{\rm exc}\,/\,S_{\rm tot}$ for H atoms agrees with the
findings by Custidiano and Jakas \cite{anti:cust05} in Fig.\
\ref{fig:Sexc_Sion}(a) for low energies and is 
still comparable for $E\le100$ keV. Their statement that $S_{\rm
  exc}\,/\,S_{\rm tot}$ is approximately a monotonously increasing function
with $E$ up to their largest $E=2.5$ MeV is also in accordance with the present
findings. For $E>300$ keV, however, their ratio lies clearly above the present
results. This is in contrast to what was observed 
for the total stopping power in Fig. \ref{fig:Eloss_H} where the results by
Custidiano and Jakas agreed with the present findings for high energies but
disagreed for low energies. Regarding these differences one may conclude that
although the total results are in agreement for high energies the underlying
physics seems not to be described correctly in one of the calculations. As has
been seen when considering the electron-energy spectra \cite{anti:luhr08a} as
well as the convergence of \aei{} high-energy electronic states become 
more important for increasing $E$. An energy cutoff for the electrons needed
in any numerical treatment has to be chosen carefully in order to obtain a
converged $S_{\rm ion}$ as shown in Fig.\ \ref{fig:Eloss_convergence} and
table \ref{tab:Eloss_convergence}. If the energy cutoff is not sufficiently
large, the \aei{} becomes too small which finally leads to a ratio $S_{\rm
  exc}\,/\,S_{\rm tot}$ that is too large. This trend was observed in the
present convergence studies.

%
%
%
\subsection{Comparison of $\bm S$ for H, H$\bm{_2}$, and He} 
\label{sec:pb_H_H2_He} 
%
%
%

\begin{figure}[t] 
    \begin{center} 
      \includegraphics[width=0.48\textwidth]{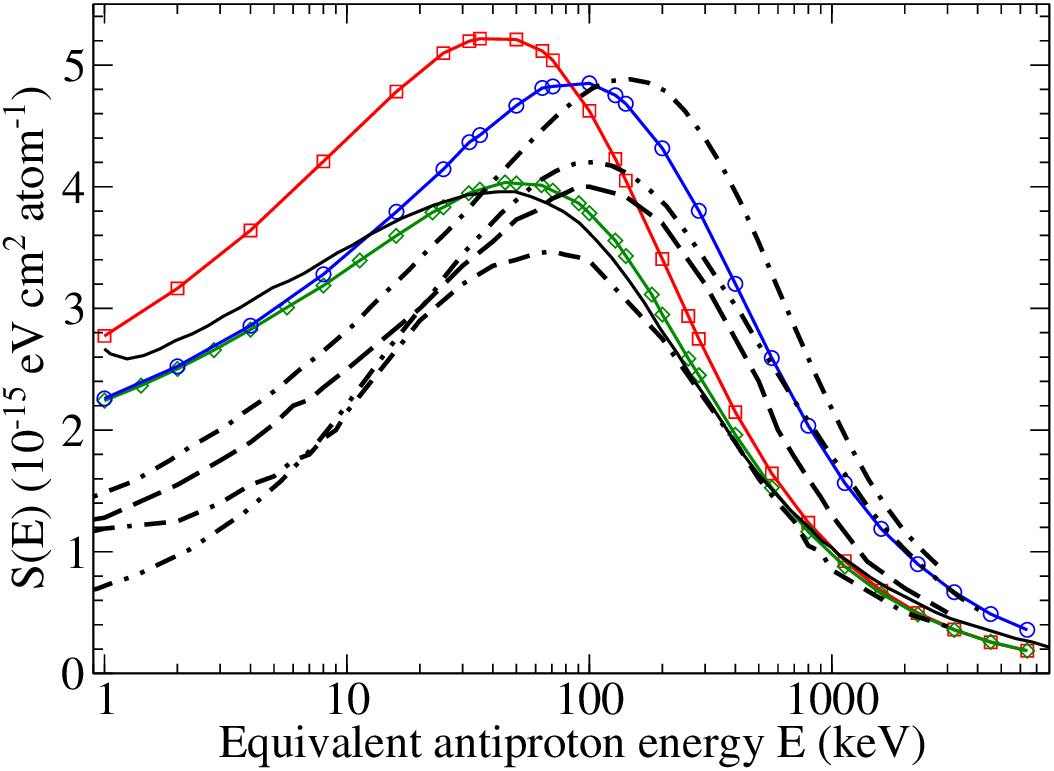} 
      \caption{(Color online) Comparison of the energy loss cross sections
        $S(E)$ for H, \hmol , and He targets as a function of the equivalent
        antiproton impact energy $E$.     
        {\bf Theory}. 
        Present results: 
        red solid curve with squares, H;
        green solid curve with diamonds, \hmol ;
        blue solid curve with circles, He.
        {\bf Experiment}.
        \hmol : 
        black solid curve, Adamo {\it et al.}~\cite{anti:adam93}, \pb ; 
        black dashed curve, Agnello {\it et al.}~\cite{anti:agne95}, \pb ; 
        black doubly-dash--dotted curve, Hauser {\it et
          al.}~\cite{sct:haus93}, \mm . 
        He:
        black dash--dotted curve, Agnello {\it et al.}~\cite{anti:agne95}, \pb;
        black dash--doubly-dotted curve, Kottmann~\cite{sct:kott87}, \mm . 
       \label{fig:Eloss_H_H2_He} }    
    \end{center} 
\end{figure} 

In Fig.\ \ref{fig:Eloss_H_H2_He} the present stopping power curves for all
three targets H, \hmol , and He are shown in one graph. For comparison also
the experimental data for \pb\ and \mm\ impact on \hmol\ and He targets are
given. In the limit of high energies $E>500$ keV the present results for
atomic and molecular hydrogen coincide which is also obvious from Table
\ref{tab:stopping_power}.  For these energies the present 
findings for He stay clearly above those for hydrogen. For $E>2000$ keV,
however, the present $S$ curve for He approaches the hydrogen results
multiplied by two (cf. Table \ref{tab:stopping_power}). 

The high-energy behavior can be made plausible by considering how the impact
parameter region --- and therefore also the distance $r$ between the electron
and the nucleus --- where the main contribution to the energy loss originates
from depends on the projectile energy. It is known that for 
large impact energies the relative importance of distant encounters for the
electronic stopping power is increasing. The used model potential in Eq.\
(\ref{eq:target_potential}) fulfills the requirement that it behaves for
$r\rightarrow \infty$ as the potential of a hydrogen atom. Since at large
distances the outer electron of the \hmol\ molecule and He atom is practically
only exposed to the field of the sum of the three remaining charges, the same
stopping power for hydrogen atoms and molecules {\it per atom} can be 
expected.  A similar argument can be applied in the case of the stopping
power of alkali-metal atoms in the case that only the valence electrons are
considered. They show the same behavior for large impact energies
\cite{anti:luhr09a} due to the fact that they also have a hydrogen-like
potential at large $r$. However, the alkali stopping power coincides with 
the hydrogen results only for higher energies $E>4000$ keV for Na, K, and Rb
and $E>1000$ keV for Li since the alkali-metal atoms are spatially more
extended than hydrogen.  

The doubled values for He for large $E$ can be
understood in the following way. For high impact velocities the collision
process can be considered independently for both electrons of the He
atom since the projectile--electron interaction happens on a much shorter time
scale than the mean electron motion and finally, the {\it dynamic}
electron-electron interaction only plays a minor role. 
Therefore, in the high-energy regime it is also meaningful to consider the
stopping power per \emph{electron} instead of per \emph{atom} leading then to
the same result for H, \hmol , and He targets.

At low energies the present stopping power curves for \hmol\ and He
coincide. 
For energies around and below the stopping maximum the results for hydrogen
atoms obviously differ from those 
for the molecules lying clearly above the He and \hmol\  curves. 
This could have been expected since all previous attempts to  
compare calculated $S$ data for H atoms with experimental curves for \hmol\
turned out to be not satisfactory. 
The apparent differences were ascribed to
{\it molecular} effects \cite{anti:schi96,anti:cabr05,anti:cust05} but were
not further specified. 
A full treatment of the \hmol\ molecule has to account for a two-center
description with two interacting electrons and vibrational and rotational
motion of the nuclei. 
This leads, e.g., to a different ionization potential
and an asymmetry of the charge distribution compared to an H atom, dynamic
two-electron effects as well as the existence of different rotational and
vibrational states.   
The present calculations employ an atomic-like one-center model for the
description of \hmol . 
It provides an appropriate ionization potential which
is, however, static since the second electron is accounted for by a screening
potential which does not allow for dynamic interaction effects. 
The nuclear motion is to a certain extent included using the linearity of the
antiproton cross sections in $R_n$. 
The present findings seem to show that the cross sections and therefore also
the stopping power are strongly determined by the correct ionization potential
of the target. Thereby, ionization is the main energy loss channel for
\pb\ collisions with \hmol . Specific molecular effects due to the existence
of two centers like rotational and vibrational motion of the nuclei,
dissociation or an asymmetric charge distribution seem to play a minor role
for impact energies above the stopping maximum. On the  other hand, dynamic 
electron-electron effects during the collision which are excluded in the
present approach seem to become important for
energies below the stopping maximum due to the fact that (i) the longer time
scales allow for interactions between the electrons and (ii) the inelastic
collisions take place closer to the nuclei where the electron density is
higher.  


%
%
%
%
\section{Conclusion} 
\label{sec:conclusion} 

Time-dependent close-coupling calculations of the electronic stopping power
for antiproton collisions with atomic and molecular hydrogen as well as helium
have  been performed in an impact-energy range from 1 keV to 6.4 MeV.  
The collision process is described using the classical trajectory
approximation.  The H, \hmol , and He targets are treated as  (effective)
one-electron systems employing a model potential which provides the correct
ground-state ionization potentials. It can be used for different fixed
internuclear distances in the case of \hmol\ and behaves like the pure Coulomb
potential of a hydrogen atom for large $r$. 

Calculations for the stopping power of hydrogen which distinguish
between atomic and molecular targets are presented and discussed considering
the existing theoretical and experimental literature, respectively. The
present He results are considered together with theoretical and experimental
results. 
The stopping power for H compares well with other non-perturbative
calculations while the He 
and \hmol\ data give a good qualitative insight but seem to overestimate $S$
for low impact energies. This might be caused by the static description of the
second electron using an effective one-electron model. For high energies the
present \hmol\ and He results agree with measurements by Adamo {\it et al.}\
\cite{anti:adam93} and for \mm\ impacts by Kottmann {\it et al.}\
\cite{sct:kott87,sct:haus93} but disagree with the findings by Agnello {\it et
  al.}\ \cite{anti:agne95} which lie above the proton stopping power.
For the highest energies $E>2$ MeV the present results for $S$ coincide for
all three targets, if the
stopping power is considered per electron instead of per atom as usually
done. This means, in the case of fast \pb\ the electrons can be
interpreted as independent particles and that the interaction takes place
mainly at large impact parameters where all electrons experience the same
potential. 

The energy loss of the projectile is for all three targets mainly caused by
ionization of electrons in contrast to alkali-metal atom targets for which
excitation is the dominant loss process. While the average energy transferred
to excited target atoms is only weakly dependent on $E$, the average energy
transferred to the ionized targets increases with $E$. Therefore, the
calculations at high energies are computationally more demanding since a basis
including high-lying continuum states is required.
 
In order to improve the description of the stopping power below the maximum a
two-electron description of the \hmol\ and He targets shall be
implemented. This would drastically reduce the uncertainties still persisting
at low impact energies and present a stringent test of the accuracy of the \pb\
measurements of the stopping power but in turn also for the ionization and
excitation cross sections of \pb\ collisions with He and \hmol .

\begin{center} 
 {\bf ACKNOWLEDGMENTS}  
\end{center} 

The authors would like to acknowledge helpful discussions with Prof.\ Knudsen
and Prof.\ Kottmann. The authors also want to thank Prof.\ Schiwietz and
Prof.\ Lodi Rizzini for correspondence. The authors are grateful to BMBF
(FLAIR Horizon) and {\it Stifterverband f\"ur die deutsche Wissenschaft} for
financial support.



\end{document}